# Two-phase water: structural evolution during freezing - thawing according to optical microscopy

T. Yakhno[1,2], V. Yakhno[1,2]

[1]Federal Research Center Institute of Applied Physics of the Russian Academy of Sciences (IAP RAS), Nizhny Novgorod, Russia;
[2]Nizhny Novgorod State University of N.I. Lobachevsky (National Research University), Nizhny Novgorod, Russia

yakhta13@gmail.com; yakhno@appl.sci-nnov.ru

**Abstract.** The structural dynamics of ice in the freezing - thawing process has been studied in the context of the concept of two-phase water. It was previously shown that water is a two-phase system consisting of free and bound (liquid crystal) fractions. Bound water is represented by hydrated shells of sodium chloride microcrystals, which are constantly present in water and are visible in the optical microscope as a dispersed phase. With rapid freezing, free water crystallizes, turning into ice with the inclusion of the dispersed phase. At this stage, xenogenic air bubbles are released. When ice melts, multiple small bubbles form from diffusely dissolved air. At the same time, the ice acquires a characteristic cellular structure. Bubbles grow, merge and form variegated air channels that contribute to the release of air into the melt water. The dispersed phase during the freezing of water under the conditions of our experiment did not show external signs of glaciation. However, when ice melted, a part of this fraction was kneaded from the main volume of water in the form of loose flakes. The loss of water of part of the microdispersed phase and most of the gases causes a change in a number of physical properties of the melt water. Possible involvement of the observed processes in the appearance of "snowballs" on the banks of rivers is discussed.

**Introduction.** The morphological features of the microstructure of fresh water and sea ice are due to water quality and environmental conditions [1,2]. In the study of the structure of ice, great attention is paid to the dynamics of the formation and release of gas bubbles accompanying the phase transitions of water as a factor determining the texture and strength of the ice cover [3-7]. The promise of a non-destructive research method, X-ray computed tomography, was shown for visualizing the processes of formation and melting of ice and frozen soils [8]. Nevertheless, laboratory modeling of these processes allows revealing some details inaccessible to direct observation in wildlife. In particular, we were unable to find in the literature descriptions concerning the observation of the morphological picture of two-phase water during freezing and thawing. This is, actually, our current work.

The concept of two-phase water arose as a means to explain a number of physical anomalies inherent to it [9-14]. To date, the evidence of the existence of high and low density water and ice has been obtained [14]. However, the areas of coexistence of these forms lie outside the natural temperatures and pressures. On the other hand, the experimental facts indicate that, under room conditions, water is a heterogeneous substance. Thus, in accordance with the NMR data, water and aqueous solutions under normal conditions are self-organizing systems that are far from thermodynamic equilibrium [15]. They form low-entropic supramolecular dissipative structures — coherent domains similar to liquid crystals. Subsequently, these results were confirmed and received an explanation in the light of a number of modern physical theories [16]. An important argument in favor of the results obtained is the experimental fact of the formation of a near-wall layer of water with anomalous properties near hydrophilic surfaces [17,18] and the further development of this topic, which led to the formation of the concept of "exclusion zone (EZ)" [19]. The structural heterogeneity of water at the micro level is also confirmed by IR spectroscopy [20], small-angle light scattering [21–

23], laser modulation-interference phase microscopy [24], precise measurement of evaporation kinetics [25] and filtering through paper filters [26]. Thus, the structural heterogeneity of water under room conditions is an irrefutable fact, despite the fact that there is no consensus as to the nature of the microphase.

In our preliminary studies it was shown that water and aqueous solutions, when observed under an optical microscope under room conditions, are microdispersed systems. The dispersion medium is unstructured at the microscopic level, and the dispersed phase is represented by a set of hydrophilic microparticles surrounded by a thick hydration shell of liquid crystal water [27,28]. In distilled water, such a structure-forming microparticle is sodium chloride microcrystal [29].

The purpose of this work was to describe the dynamics of microstructural transformations of water during freezing - thawing under laboratory conditions, taking into account the dispersion of the medium under study.

**1. Materials and methods**

The work was performed under the natural conditions of the laboratory: T = 21-23°C, H ~ 67-70%. As the material of the study we used tap and distilled water, as well as ice formed during the hardening of these waters. Part of the ice samples were prepared from boiled water. Water was microscoped in a thin layer, between the slide and cover glass, before and after freezing in the chamber of the domestic cooler (T = -20°C). The thickness of the layer was about 18 μm. Other ice samples were prepared as follows. Water was poured into clean transparent-plastic Petri dishes with a layer of ~ 2–3 mm and placed in a freezer (T = -20° C) for at least 30 min (Fig. 1).

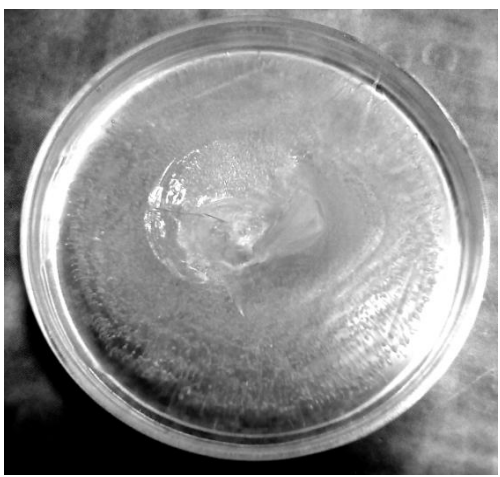

**Fig. 1.** The object to be examined is ice in a Petri dish.

After the formation of ice, the sample was immediately transferred under a Levenhuk microscope with a Levenhuk C-1400 video camera interfaced with a computer using the ToupView program, which allows for photo and video recording. Slides and coverslips produced by ApexLab (Russia) with dimensions (25.4 x 72.2 mm x 1 mm) and (24 x 24 x 0.6 mm), respectively, Petri dishes d = 35 mm (polystyrene, sterile, MiniMed, Russia) were used in our work. Part of the water samples were frozen on a glass slide without a coverslip. Observations were carried out before the complete transition of ice into the liquid phase. The bottom of the Petri dish was also investigated after evaporation of the frozen - thawed water.

## 2. Results and discussion

The surface of the ice made from tap water was completely covered with icy air bubbles (Fig. 2a). In a sample of ice made from boiled tap water, the number of such bubbles was much smaller (Fig. 2b).

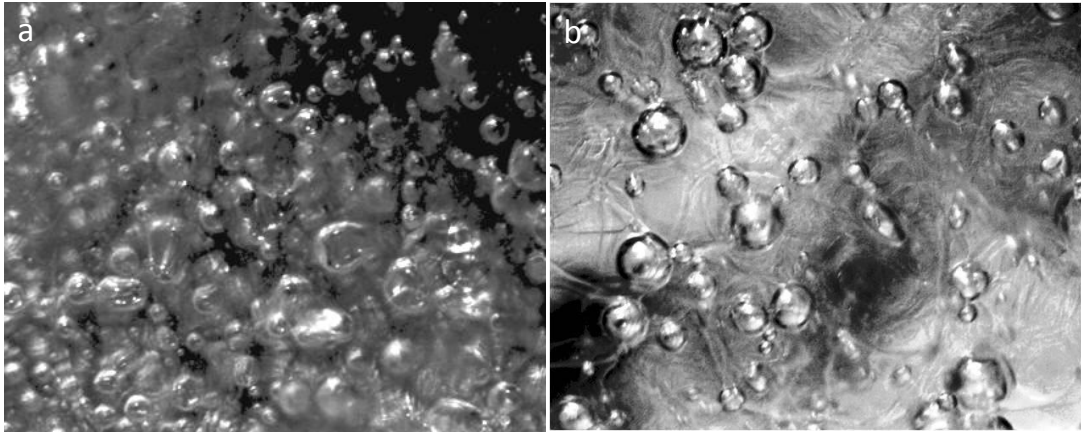

**Fig. 2.** The surface of ice made from tap water: a - unboiled; b - boiled. The width of each frame is 3 mm.

The surface of ice from distilled water looked smooth and homogeneous, with rare inclusions of bubbles that were smaller in boiled water (Fig. 3, a, b).

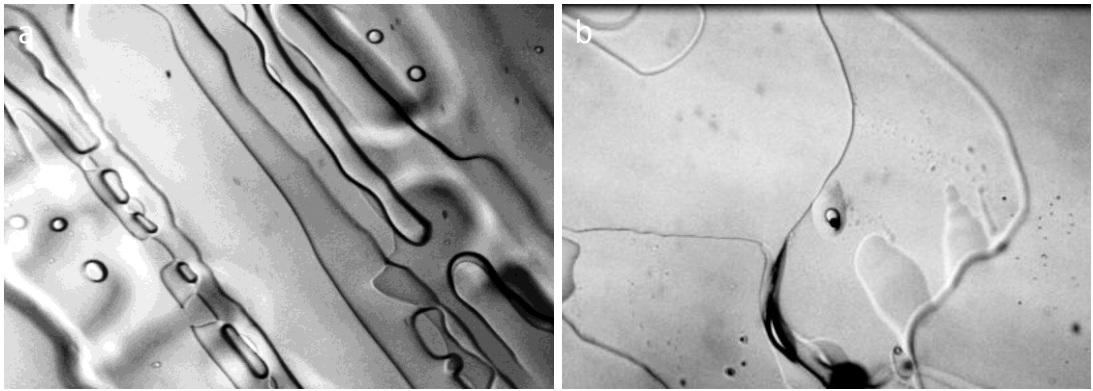

**Fig. 3.** The surface of ice made from distilled water: a - unboiled; b - boiled. The width of each frame is 3 mm.

In our case, the tap water contains more air impurities, because it was under the conditions of high pressure in the water supply system (4 atm) for a long time. Boiling led to a significant decrease in the concentration of "impurity" air in the water. According to the data of [1], the total number of xenogenic air inclusions can reach 50% of the ice volume in natural fresh ice. That is, a large part of air bubbles contained in ice is a product of air capture by water as a result of disturbances at the boundary with the atmosphere and the evolution of gases from the bottom of reservoirs. This xenogenic air during rapid freezing is forced out at the interface and forms a layer of icy bubbles. As it melts, homogeneous ice acquires a cellular structure due to the release of bubbles of diffusely dissolved air (Fig. 4).

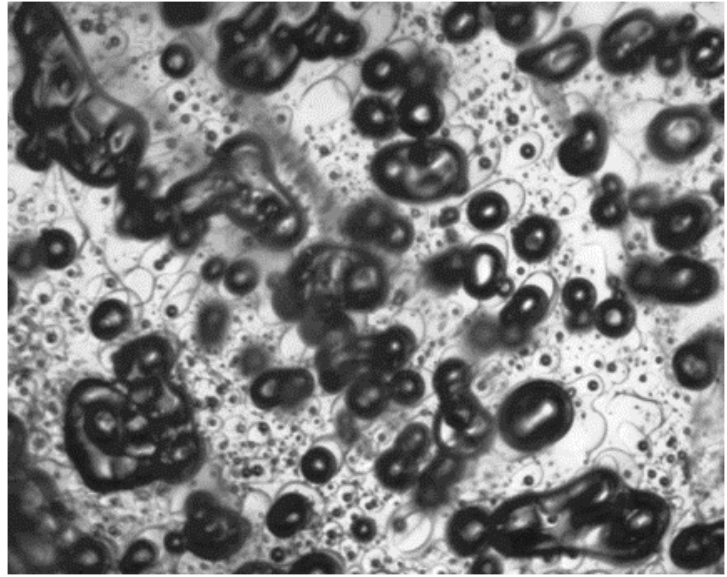

**Fig. 4.** The cellular structure of ice in the allocation of dissolved air. Large bubbles are formed on the upper surface. Frame width - 3 mm.

The release of air is accompanied by compaction of the surrounding ice, and a recess is formed around each growing bubble. As air is released, these depressions in the thickness of the ice grow and merge, forming different-sized air passages (Fig. 5).

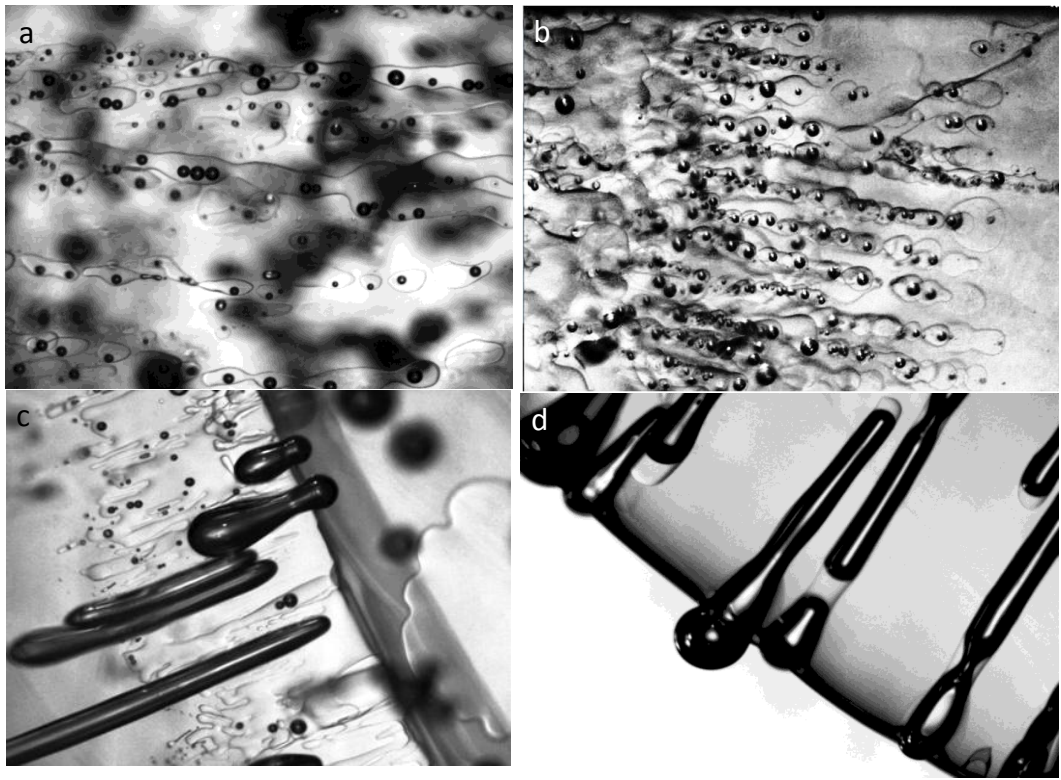

**Fig. 5.** The formation of air bubbles from the air dissolved in ice and the formation of air passages as they grow (a-d). The width of each frame is 3 mm.

The process of melting the plates of homogeneous ice is accompanied by the release of air bubbles into liquid water (Fig. 6).

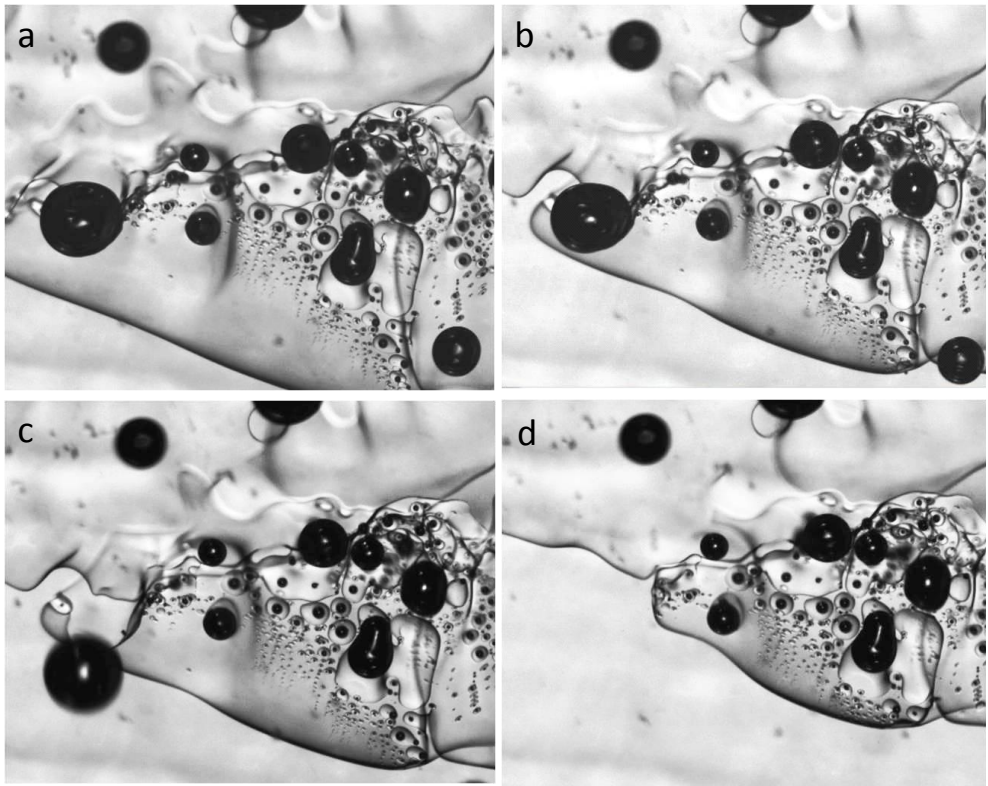

**Fig. 6.** Release of air bubbles from homogeneous ice into liquid water as it melts. Sequential snapshots (a-d). The width of each frame is 3 mm.

    The solubility of air in water at a temperature of + 20°C and a pressure of 1 atm is 19 ml / l of water [30], and at 0°C and the same pressure - 29.18 ml / l of water [31]. That is, when water is cooled from room temperature to 0°C, the air solubility in it rises by more than 50%. However, with the formation of ice, some of the air is released. This is mainly xenogenic air, trapped in water from the outside. When ice melts, air bubbles are also released. But this is diffuse air dissolved in water. Bubbles originate in a layer of homogeneous ice, grow to form cavities in the ice, passing into channels, and, ultimately, enter the liquid phase of thawed water.

    Due to the high light scattering ability of the ice surface, the dispersed phase of water is usually invisible. Figure 7 shows a fragment of a drop of distilled water with a volume of 1 ml frozen on the surface of a glass slide. In the inset on the left, the surface of the ice looks homogeneous. However, when the image fragment is enlarged 6 times, the structure becomes noticeable, namely, the accumulation of the dispersed phase immobilized in homogeneous ice.

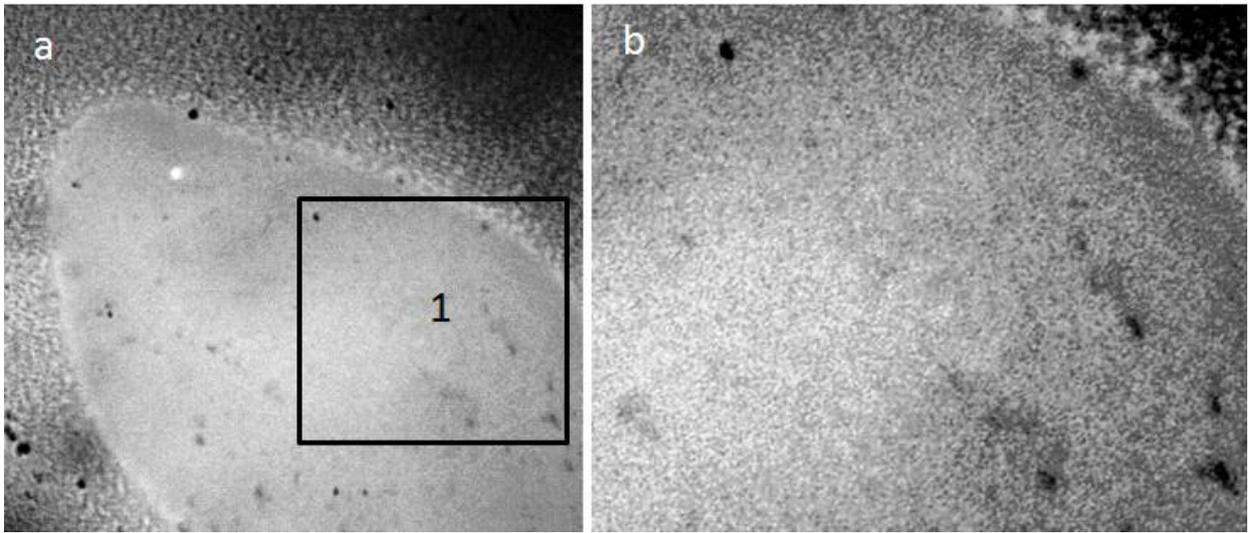

**Fig. 7.** A drop of distilled water, frozen on the surface of a glass slide: on the left - a frame width of 2.5 mm; on the right - a fragment of image 1, enlarged 6 times.

The structure of ice, the dynamic and morphological features of its melting depend on temperature conditions, freezing rate and volume limitation [32]. We examined a drop of water during a quick freeze in the open air. Now let's see how a thin layer of water freezes in conditions of limiting its volume from above and below with slide and coverslip. The layer thickness is ~ 18 μm (Fig. 8).

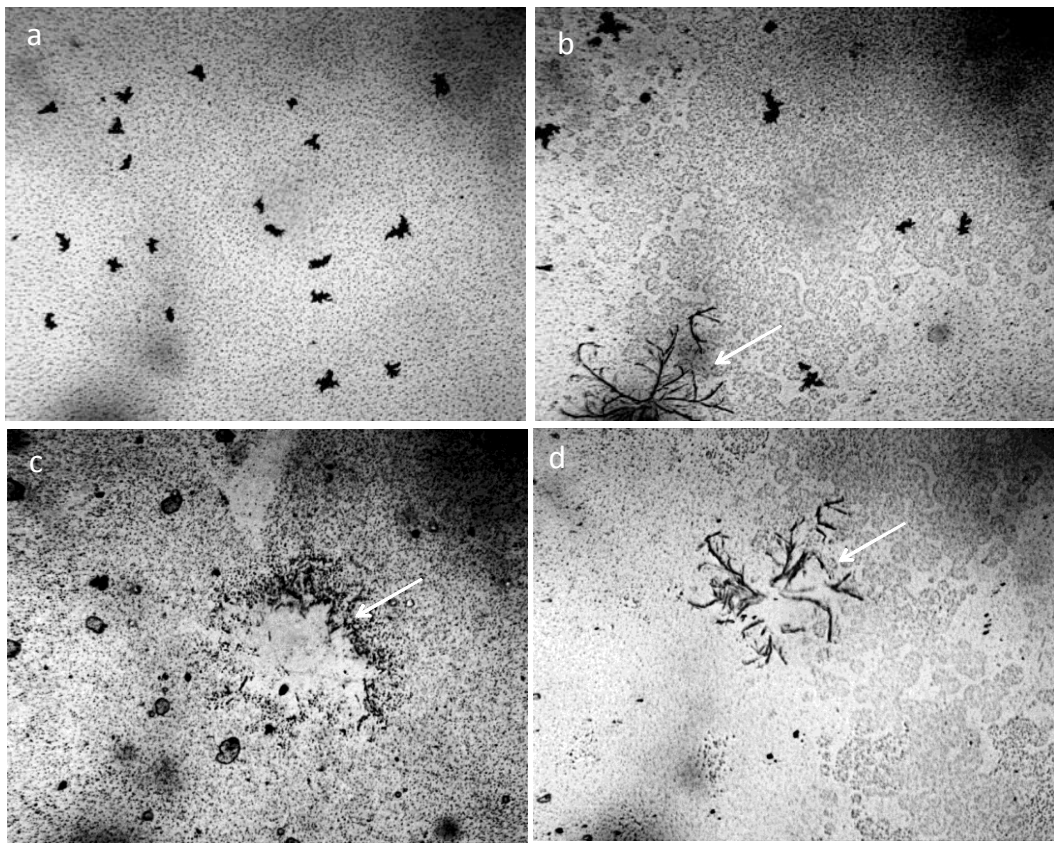

**Fig. 8.** A thin layer of water enclosed between the slide and the coverslip: (a, b - before freezing the drug, different fields of view; c, d - after freezing the drug, different fields of view). Arrows point to spherulites. Black fractal clusters — hydrophobic trace impurities — are common inclusions observed in distilled water.

It can be seen that the appearance of the dispersed phase during the freezing of water in a thin layer bounded by glasses at the top and bottom practically did not change (Fig. 8). At the same time, the spherulite during freezing is covered with amorphous ice formed from structureless (at a given level of scale) water. It is known that the heat of crystallization of bound water is greater than the heat of crystallization of free water [33], that is, the dispersion medium should freeze earlier than the dispersed phase, which corresponds to our observations. When ice is thawed in a Petri dish, part of the dispersed phase is stirred in the form of aggregates or large flakes (Fig. 9). Apparently, this process is associated with partial damage to the interfacial surfaces of the liquid crystal spheres during freezing and strengthening the adhesive contacts between them.

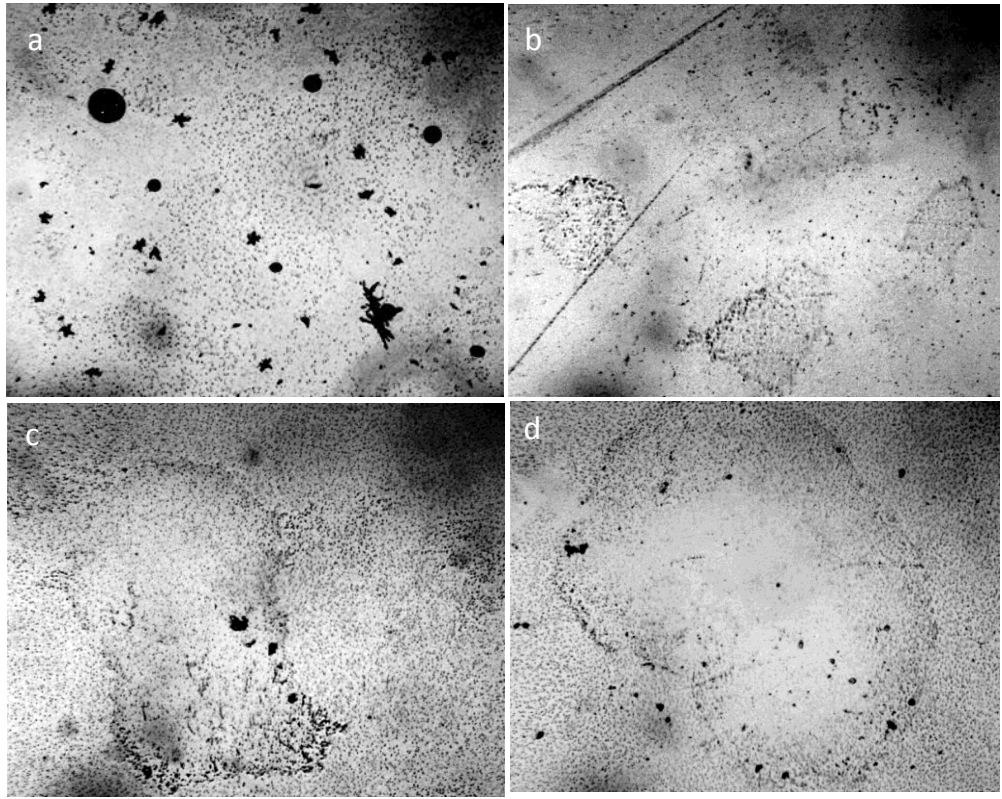

**Fig. 9.** Large aggregates of the dispersed phase floating in the melt water. The width of each frame is 3 mm.

Note that in the preparation of "melt water" at home, this fraction, along with other micronutrients, is likely to fall into the "pickle" and will be drained. We will do the same and see what remains in the melt water after draining the thawed portion of water (~ 50%) and evaporation of residues of the liquid phase.

Three days after evaporation of water, wet drops of liquid could be observed on a dry bottom of a Petri dish (Fig. 10). When the focal length inside each drop was changed, growing NaCl microcrystals were detected. This phenomenon was described by us earlier [29].

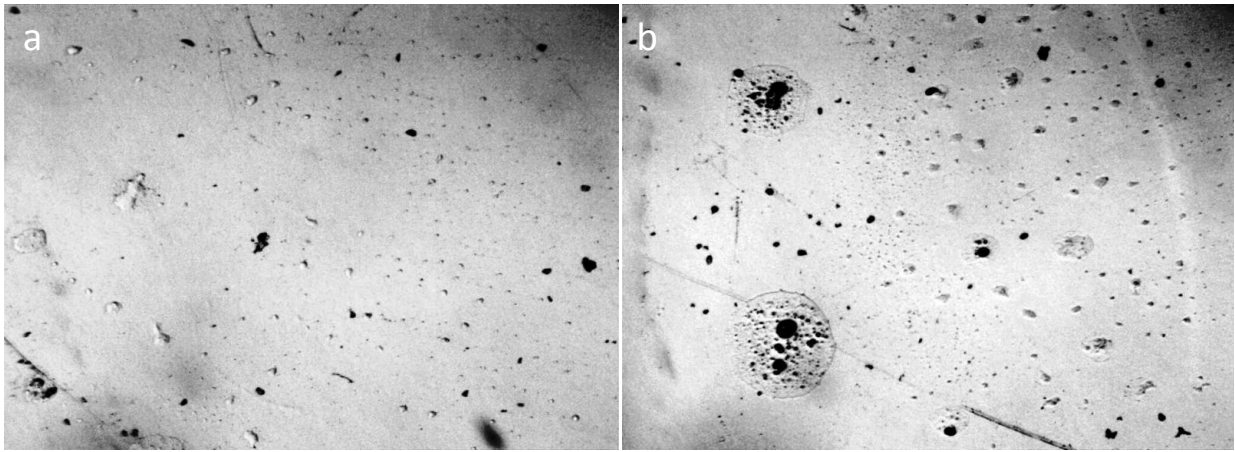

**Fig. 10.** Sediment on the bottom of a Petri dish 3 days after evaporation of the residual melt water: a - undried drops; b - on the left - aggregates of the dispersed phase with salt crystals growing in them, on the right - small aggregates of the dispersed phase in a liquid shell with salt crystals. In the center of the frame is a monolayer of aggregated particles of the dispersed phase. The width of each frame is 3 mm.

We have previously shown that in the liquid phase, the hydration shell of the central hydrophilic particle is similar to a viscous liquid, it does not evaporate at room temperature, but evaporates at a temperature of about 300°C [34]. This hydration shell undergoes erosion and collapses with increasing ionic strength (osmotic pressure) of the solution [27,28]. These physical indicators increase, in particular, when free water evaporates.

Let us see what the microstructure of water looks like in a thin layer between the slide and the coverslip before and after the freeze-thaw procedure in a Petri dish (Fig. 11).

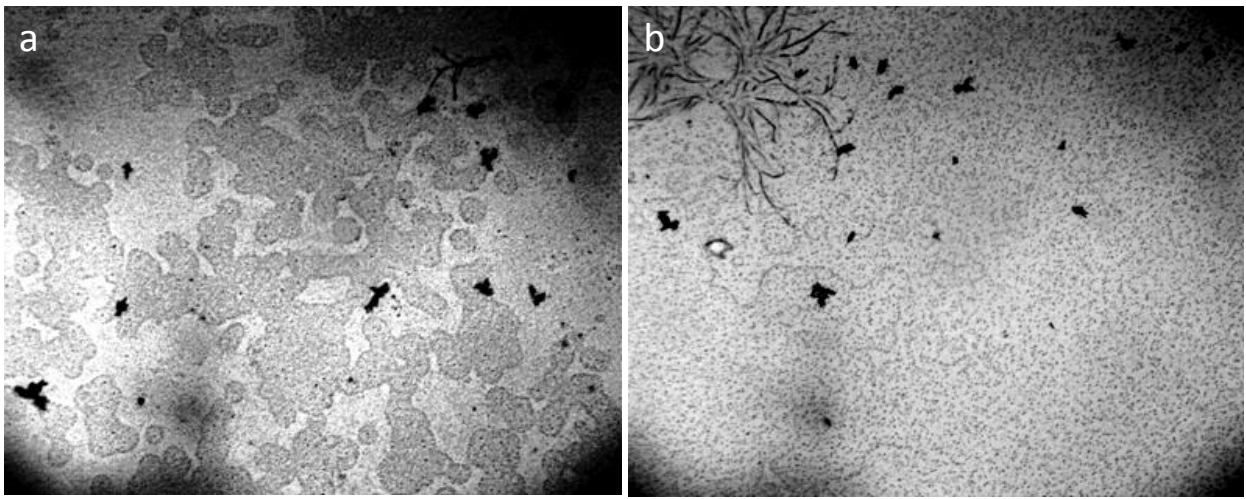

**Fig. 11.** The microstructure of a thin layer of water before (a) and after (b) freezing and thawing of water in the volume of a Petri dish. The width of each frame is 3 mm.

It is obvious that the nature of clustering of the dispersed phase has changed; the optical density of the sample has decreased. After freezing and thawing, the microstructure of the water looks friable. In accordance with our observations, the melt water is released from part of the dispersed phase which is kneaded in the form of flakes, and from the main mass of gases. This, in our opinion, causes changes in a number of physical properties of melt water, in particular, thermal conductivity [35]. It is interesting to compare the data we obtained with the results of the work [36], in which during the melting of ice in the sample under study, absorption in the 270 nm band, specific to the exclusion zone (EZ), appeared and soon disappeared.

The observations of the dynamics of ice melting allow us to put forward an idea about the unexplained "bottom ice" phenomenon. We speak about the occurrence of a large number of ice balls carried by the waves on the banks of unfrozen rivers sometimes in the winter (Fig. 12).

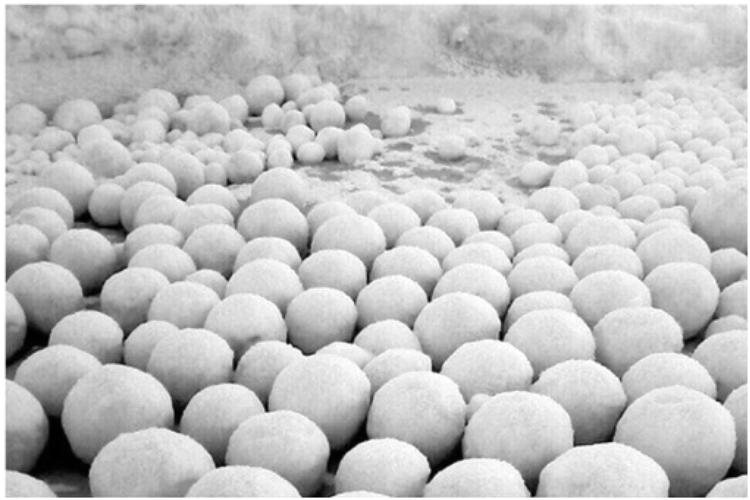

**Fig. 12.** Ice balls on the Ob' river [39].

This question was thoroughly studied by the Russian scientist V.Ya. Altberg in the first half of the last century [37,38]. In particular, it was shown that the formation of balls usually occurs at sub-zero air temperature, when the river does not have a solid ice cover. As a result, the water in the river is supercooled (not more than 0.1 ° C), and the conditions for the growth of crystalline nuclei inside the liquid are created. We could see that the freezing water displaces the air first in the form of bubbles. The bubbles immediately turn into icy spheres (hollow inside). These areas may well serve as a "seed" for fouling crystals and the formation of snow balls. The mechanism of such growth is possible, and is described for the round head of a cryosurgical instrument [40].

## 3. Conclusion

The study showed that, in our experiment, the structureless water and the microdispersed phase during freezing and thawing behave in different ways. No pronounced signs of glaciation of the dispersed phase were observed under the conditions of our experiment, while the dispersion medium exhibited the known morphological properties of ordinary ice. In the process of thawing, part of the microdispersed phase formed loose flakes, "kneading" from the liquid phase, which may be the result of surface temperature damage of a part of the liquid crystal spheres. The release of gas bubbles during freezing and thawing of water occurs in two stages: 1) when ice forms, bubbles of xenogenic "trapped" air are released, and 2) when ice melts, bubbles of endogenous, diffusely dissolved air are released. The path of formation of air channels in the thickness of the ice was traced also. Observation of the nature and dynamics of the precipitate after evaporation of melt water confirmed our understanding of the physical nature of the microdispersed phase: these are microcrystals of salt — sodium chloride covered with a thick layer of hydrated liquid-crystal water. The loss of water of part of the microdispersed phase and most of the gases causes changes in a number of physical properties of the melt water. Based on the observations, it was assumed that the origin of the considered natural phenomenon is the appearance of ice balls on the banks of rivers the appearance

We hope that our observations of the behavior of two-phase water during freezing and thawing will be useful for further glaciological studies.

**Acknowledgments**

The work was supported by the Ministry of Education and Science of Russia (Project No. 14.Y26.31.0022).

**References**

1. Tyshko K.P., Cherepanov N.V., Fedotov V.I. Crystal structure of sea ice cover. StPb: Gidrometizdat, 2000, 66 p. (In Russian).
2. Bezrukov Yu.F. Oceanology. Part 1. Physical phenomena and processes in the ocean. Simferopol, 2006, 162 p. https://www.twirpx.com/file/282007 (In Russian).
3. Maeno N. Air Bubble Formation in Ice Crystals. In Physics of Snow and Ice: proceedings Hokkaido University Collection of Scholarly and Academic Papers, 1967, 207–218.
4. Hruba J., Kletetschka G. Environmental record of layers of bubbles in natural pond ice. // Journal of Glaciology, 2018, 1-11. doi: 10.1017/jog.2018.73.
5. Bari S.A., Hallett J. Nucleation and growth of bubbles at an ice-water interface. // Journal of Glaciology, 1974, 13(69), 489–520.
6. Yoshimura K., Inada T. and Koyama S. Growth of spherical and cylindrical oxygen bubbles at an ice-water interface. // Cryst. Growth Des., 2008, 8(7), 2108–2115.
7. Zhekamukhov M.K. Distribution of dissolved gas in water and bubbles in ice upon movement of a crystallization front. // J. Eng. Phy., (1976) 31(4), 1158–1162.
8. Romanenko K.A., Abrosimov K.N., Kurchatova A.N., Rogov V.V. Experience in the use of X-ray computed tomography in the study of microstructure of frozen rocks and soils. // Earth's Cryosphere, 2017, 21(4), 75–81. (In Russian).
9. Maestro, L.M., Marqués, M.I., Camarillo, E., Jaque, D., García Solé, J., Gonzalo, J.A., Jaque, F., del Valle, J.C., Mallamace, F. and Stanley, H.E. On the existence of two states in liquid water: impact on biological and nanoscopic systems. // Int. J. Nanotechnol., 2016, 13(8/9), 667–677.
10. Gallo P., Amann-Winkel K., Angell C.A., Anisimov M.A., Caupin F., Chakravarty C., Lascaris E., Loerting T., Panagiotopoulos A.Z., Russo J., Sellberg J.A., Stanley H.E., Tanaka H., Vega C., Xu L., Pettersson L.G. Water: A Tale of Two Liquids. // Chem. Rev. 2016, 116, 7463−7500.
11. Stanley, H. E.; Kumar, P.; Franzese, G.; Xu, L.; Yan, Z.; Mazza, M. G.; Buldyrev, S. V.; Chen, S.-H.; Mallamace, F. Liquid Polyamorphism: Possible Relation to the Anomalous Behavior of Water. // Eur. Phys. J.: Spec. Top. 2008, 161, 1−17.
12. Moore E.B., Molinero V. Structural transformation in supercooled water controls the crystallization rate of ice. Nature, 2011, 479(7374), 506–508.
13. Fuentevilla D. A., Anisimov M. A. Scaled equation of state for supercooled water near the liquid-liquid critical point. // Phys. Rev. Lett. 2006, 97, 195702.
14. Perakis F., Amann-Winkel K., Lehmkühler F., Sprung M., Mariedahl D., Sellberg J. A., Pathak H., Späh A., Cavalca F., Schlesinger D., Ricci A., Jain A., Massani B., Aubree F.,. Benmore C. J, Loerting T., Grübel G., Pettersson L. G. M.,  Nilsson A.. Diffusive dynamics during the high-to-low density transition in amorphous ice. // PNAS, 2017, 114(31), 8193–8198.

15. Marchettini N., Niccolucci V., Tiezzi E. The supramolecular structure of water. // from: Designe and Nature, C.A. Brebbia, L. Sucharov & P. Pascola (Editors). ISBN 1-85312-901-1. www.witpress.com
16. Tiezzi E., Catalucci M., Marchettini N. The supramolecular structure of water: NMR studies. // Int. J. of Designe & Nature and Ecodynamics. 2010, 5(1), 10-20.
17. Deryagin B.V., Churaev N.V., Fedyakin N.N., Talaev M.V., Ershova I.G. The modified state of water and other liquids. Izv. Academy of Sciences of the USSR, a series of chemical, 1967, No. 10, 2178. (In Russian).
18. Deryagin B.V. New data on superdense water. Uspekhi Fizicheskikh Nauk, 1970, 100(4) 726-728. (In Russian).
19. Pollack G.H. Cells, gells, and the engines of life. Ebner & Sons, Seattle. WA USA, 2001, 294 p.
20. Fesenko E.E., Terpugov E.L. On the unusual spectral properties of water in a thin layer. // Biofizika, 1999, 44(1), 5-9. (In Russian).
21. Sedlák M. Large-scale supramolecular structure in solutions of low molar mass compounds and mixtures of liquids: I. Light scattering characterization. J. Phys. Chem. B. 2006, 110(9), 4329–4338.
22. Smirnov A.N., Lapshin V.B., Balyshev A.V., et al. The structure of water: giant heterophase clusters of water. // Chemistry and technology of water, 2005, № 2, p. 11-37. (In Russian).
23. Bunkin N.F., Shkirin A.V., Kozlov V.A., Starosvetskij A.V. Laser scattering in water and aqueous solutions of salts. // Proc. of SPIE, 2010, 7376, Article Number: 73761D.N.F.
24. Bunkin N. F., Shkirin A. V., Kozlov V. A., Starosvetsky A. V., Ignatiev P. S. Quasistable clusters of dissolved gas nanobubbles in water and aqueous electrolyte solutions. // Nanosystems, Nanomaterials, Nanotechnologies, 2011, 9(2), 499-504.
25. Novikov S.N., Ermolaeva A.I., Timoshenkov S.P., Minaev V.S. The influence of supramolecular structure of water on the kinetics of vaporization. // Russian Journal of Physical Chemistry A, 2010, 84(4), 534-537.
26. Conti A., Palombo M., Parmentier A., Poggi G., Baglioni P., De Luca F. Two-phase water model in the cellulose network of paper. // Cellulose, 2017, 24, 3479–3487. DOI 10.1007/s10570-017-1338-2.
27. Yakhno T.A., Yakhno V.G. Water-induced self-oscillatory processes in colloidal systems by the example of instant coffee. // Journal of Basic and Applied Research International, 2017, 20(2), 70-83.
28. Yakhno T.A., Yakhno V.G.. A study of structural organization of water and aqueous solutions by means of optical microscopy. https://arxiv.org/ftp/arxiv/papers/1809/1809.00906.pdf
29. Yakhno T., Drozdov M., Yakhno V. Giant water clusters: where are they from? https://arxiv.org/ftp/arxiv/papers/1810/1810.05452.pdf
30. Polling L., Polling P. Chemistry. M., 1978, p. 263. (In Russian).
31. Chemist Handbook, T.5, L.: Chemistry, 1968, p. 25. (In Russian).
32. Sigunov Yu.A., Samylova Yu.A. Dynamics of pressure growth during freezing of a closed volume of water with a dissolved gas. // Applied mechanics and technical physics, 2006, 47 (6), 85-92. (In Russian).
33. Starostin E.G., Lebedev M.P. Properties of bound water in dispersed rocks. Part II. The heat of crystallization. // Earth's Cryosphere, 2014, 18 (4), 39-46. (In Russian).
34. Yakhno T.A., Yakhno V.G., Zanozina V.F. Phase transitions of water as a source of slow oscillatory processes in liquid media. // Proceedings of XII International Science-Technical

Conference "Modern trends in biological physics and chemistry", Sevastopol, 2-6 of October, 2017, 23-27 (in Russian).
35. Boshenyatov B.V. The role of the interaction of particles in the cluster model of thermal conductivity of nanofluids. // Letters to the Journal of Applied Physics, 2018, 44 (3), 17-24. (In Russian).
36. So E., Stahlberg R., Pollack G. H. Exclusion zone as intermediate between ice and water. WIT Transactions on Ecology and the Environment, 2011, 153. WIT Press. ISSN 1743-3541 (on-line). doi:10.2495/WS110011. https://www.witpress.com/elibrary
37. Altberg V.Ya. On the bottom ice (a report read at the congress on plumbing in Moscow, in October 1922). // Advances in Physical Sciences, 1923, 3, 392–403. (In Russian).
38. Altberg V.Ya. To the problem of nucleation of crystals. // Advances in Physical Sciences, 1939, 21 (1), 69-74. (In Russian).
39. Ice balls on the Ob', according Altapress. http://altapress.ru/story/issledovateli-rasskazali-o-proishozhdenii-gigantskih-ledyanih-sharov-na-obi-190120.
40. Kapreljants A.S., Migunova R.G., Kapreljants A.A. Cryomicroscopic study of the formation of ice crystals around a cryosurgical probe. // Problems of cryobiology, 1999, 2, 70-75. (In Russian).